# Image Convolution-Based Method and Statistical Modelling for Experimental Height Estimations of Laminar to Transition-to-Turbulent Regime Diffusion Flames

M. De La Cruz-Ávila[a]*, J.E. De León-Ruiz[b]*, I. Carvajal-Mariscal[b], F. Peña-Polo[b, c], L. Di G. Sigalotti[d]

[a] Universidad Nacional Autónoma de México, Instituto de Ingeniería, Ciudad Universitaria, CDMX, 04510, México.
[b] Instituto Politécnico Nacional, ESIME – UPALM, CDMX, 07738, México.
[c] Centro de Física, Instituto Venezolano de Investigaciones Científicas, IVIC, Caracas, 20632, Venezuela.
[d] Universidad Autónoma Metropolitana-Azcapotzalco (UAM-A), Departamento de Ciencias Básicas, Av. San Pablo 180, 02200 CDMX, México.
**\*Correspondence:*mauriciodlca1@gmail.com; jedeleonr@gmail.com

**ABSTRACT**

This paper presents an experimental methodology to measure the height of the flame using convolution image processing and statistical analysis. The experimental setup employs a burner with four circularly arranged nozzles. Six different volumetric fuel flows were employed, and flame images were captured from three different visualization planes utilizing a three high-definition camera array, a thermal imaging camera and an image-processing algorithm. The flame height was indirectly measured using pixel quantification and conversion through a reference length. Although the fuel flow was the most significant factor, the visualization plane and the image source were also found to be particularly relevant, since certain flame features were only perceivable depending on the approach. The measurements were compared to different existing theoretical correlations, yielding an overall adjustment ranging from 3.25 to 3.97cm. The present methodology yields an overall statistical tolerance of 1.27 cm and an expanded uncertainty of 0.599 cm. Furthermore, the thermal imaging has revealed a consistent difference in the overall luminous observable flame of 2.54 cm. For this particular burner configuration, correlations were derived by statistical modelling, which explain the flame height fluctuations with an average setting of 97.23%.

## I. INTRODUCTION

In combustion science, a proper description of the diffusion flame phenomena requires accurate measurements in order to obtain relevant and representative data about the several features that define the flame front development and structure. These include the flame front length, the flame structure luminescence, the stoichiometric mixing zone and the expansion flame radius, among other features, as has been revealed by numerical [1–4] and experimental studies of diffusion flames [5–8]. However, the length measurement, associated to the overall flame development, also involves certain features that need to be considered beforehand as, for example, the emergence of several inherent layers, each with their own distinct features and corresponding lengths, such as the luminous flame, the continuous and intermittent flame, the chemical flame and the flame core, to name a few [9–14].

For many past decades there has been a strong interest to explain and predict the flame lengths. A major review on this subject is given by Becker and Liang [15], where many definitions for the flame length can be found in the references therein. However, none of these definitions is preferred to the others. Therefore, care must be exercised in comparing results of different authors and in the application of correlation formulae. Common definitions of flame length include visual determinations as obtained by averaging several individual instantaneous visible flame lengths from photographic records.

These definitions are determined by the greatest vertical distance, along the flame centreline, between the burner nozzle base and (i) the tip of the visible flame, (ii) the location of the average peak centreline temperature and (iii) the location where the fuel and oxidant reach stoichiometric concentration;

the latter distance being the most widely accepted [16]. Based on these different interpretations, care must be taken when comparing results from different investigations and using their proposed empirical correlations.

Many efforts have been done to consolidate different theories and correlations as well as diverse data [17–22]. In particular, Delichatsios [23], Reshetnikov and Frolov [24], Blake and McDonald [25] and Heskestad [26,27] have proposed general wide-range functions to determine the flame height; mainly for fuel pool fire as well as premixed and co-flow diffusion flames. However, for recent inverse diffusion flames or port array configurations, neither low nor high pressures have been tested due to the buoyant-flickering nature of the flame.

The visible length of a non-premixed flame is still an important indicator of both (i) the overall fuel-oxidant mixing process, since the flame length is proportional to the axial distance required to dilute the fuel mixture fraction to its stoichiometric value [28], and (ii) the flame structure model as well as the ash-particles residence-times [16]. Because of this, Zukoski et al. [29] proposed for the first time a correlation using more extensively the Froude number to determine a general flame length, based on heat release, gravity effects and density gradients, which was later on adopted by Delichatsios [23]. In general, flame heights have been measured from videotape recordings and by optical averaged techniques, whose output albeit appropriate, was not very accurate. Moreover, visible flame lengths tend to be greater than those based on temperature or concentration measurements. For example, Turns and Bandaru [30] reported temperature-based flame lengths which were approximately from 65 to 80% larger than time-averaged visible flame lengths, depending on the fuel type.

The knowledge of the diffusion flame length is important for both practical and conceptual reasons. In particular, an increased flame height can result in an enhanced thermal NOx production [25,31,32]. In theory, the visible length of a non-premixed flame is an important indicator of the overall fuel-oxidant mixing process, since the flame length is proportional to the axial distance required to dilute the fuel mixture fraction to its stoichiometric value [28]. Flame height measurements have been used to test flame structure models and calculate residence times of ash particles [16]. In the industry, the flame length is of particular interest because, appropriate separation distances between the injection equipment and the burner walls have to be specified for a given flame length [3], which is deemed to be the base of safety considerations for any fire.

The most commonly accepted definition of flame height is given by the distance from the burner to the position in the centreline where the fuel and the oxidant are in stoichiometric proportions [16]. A more accurate and less subjective methodology based on measurements of species along the flame axis is proposed here. This methodology is also suitable for scenarios where observations of the luminous structure height of the flame are not possible [13].Therefore, from knowledge of these characteristics, it would be possible to design, develop and enhance combustion chambers prototypes regarding the nature of the diffusion flame phenomenon as well as to improve and optimize the technology for in-situ steam-generator, spacecraft propulsion and pyrolysis reactor combustors, just to mention a few.

The aim of the present work is: (i) to measure an average height given solely by the nature of the flame and not by a mathematical prediction model, (ii) derive a flame height correlation from the measurements made, (iii) find if the oscillation and its observable concomitant phenomena have a significant effect on the resulting measurement of height and (iv) based on the output yielded, propose the present experimental procedure as an alternative methodology for flame length measurements.

## II.     COMBUSTION PHENOMENA: DIFFUSION FLAME PARTICULARITIES

In order to properly design and conduct a measurement procedure, the basic nature of the flame must be accounted for in advance. Therefore, it will be necessary to understand the combustion phenomena as well as the fundamental factors that determine its magnitude.

### A. Flame Height Definition

The most commonly accepted definition of flame height is the average distance measured vertically, up and down from the base of the burner rim up to the position on the centreline (tip of the flame) where the fuel and oxidant are in stoichiometric proportions [16,33].

For cylindrical jet diffusion flames, the height, $h_{Lf}\big|_x$, is defined as the difference between two points along the burner axis [33]. However, when the fuel is supplied through a circular nozzle the fuel concentration gradient, $dY_{fuel}/dr$, which enhances the diffusion and velocity of the fuel, $u_{fuel}$, exerts no significant effect on the resulting flame height, thereby rendering the buoyancy force negligible. Furthermore, it is considered that the burning process is not affected by the mixing rate, based on the assumption that the air and fuel jet react immediately when they are exposed to each other.

Moreover, based on the mass diffusivity equation for gases at a constant fuel mass flow, the flame height is considered to be independent of the pressure. In addition, the flame height is also independent of the gas-burner diameter for a particular volumetric fuel flow, $\dot{V}_{fuel}$. Therefore, the flame height will only depend on the fuel mass flow, $\dot{m}_{fuel}$, Despite being proportional to the fuel mass flow, the resulting magnitude of the diffusion flame height is not seen to depend linearly on the fuel mass flow.

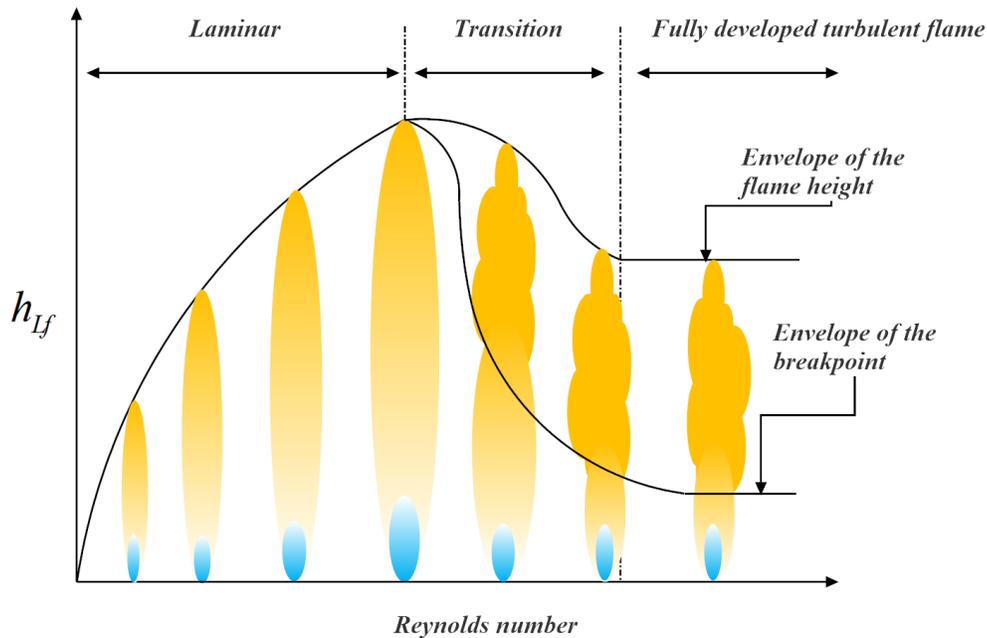

Fig. 1: Variation of the flame height with the Reynolds number in a diffusion flame [34].

For a given fuel and burner, the flame height exhibits steady increments with increased jet velocity in the laminar regime. Beginning at a low flow rate, the flame height increases with increasing flow rate, reaches a maximum, and then shortens. Shortly before the maximum height is reached, the flame begins to flicker (i.e., it becomes turbulent) at the top. This leads to a sudden and continuous decrease of the flame height accompanied by an outward spreading of the flame boundary.

This occurrence separates "laminar diffusion" from flames of "transitional" type. With further increase of the flow rate, the flickering spreads in a downward direction. The spreading stops rather suddenly when only a few diameters above the port are left free of fluctuations. At this point, the flame height becomes independent of the flow rate, and the "transition flame" becomes the "turbulent flame" (see Fig. 1.), which produces more noise, less soot, and reduced luminosity [35]. However, at a fuel-jet critical-Reynolds number, the flame is lifted from the burner rims and as the jet velocity increases, the flame lifts even higher, becoming gradually smaller until it is blown off completely. This point is known as the blow-off limit and depends on the fuel-oxidizer system involved [36,37].

Eventually, the flickering phenomena ensue in the flame, which is related to its stability. This characteristic reveals the vibratory nature of the luminous flame, particularly at the flame-end tip. The flickering frequency is proportional to both the nozzle diameter and the fuel jet velocity. This could be categorized into two different types: (i) bulk flickering and (ii) tip flickering [38].

The bulk flickering ensues at sufficiently low Reynolds numbers and is caused by the buoyant-flow instability driven by large-scale toroidal vortices, which then leads to a fire-plume periodic separation from the top of the main flame structure [23,38–43]. In contrast, the tip flickering occurs at high Reynolds numbers and it is mainly due to the strong shear force around the fuel jet, which then results in an oscillation and elongation of the top of the flame [38]. It is worth noting that flickering is indeed quite difficult to measure due to its random nature. However, its effects are easily noticeable due to the presence of flame-tip tilt/crookedness, uneven kernel formation and/or plume detachment.

From the previous considerations, the determination of the luminous flame height, $h_{Lf}|_x$, is given in terms of empirical correlations based on both the continuous, $h_{cf}|_x$, and intermittent flame correlations, $h_{if}|_x$, which measure the height of the flame body and quantify its variation due to flickering, respectively, where

$$h_{Lf}|_x = h_{cf}|_x + h_{if}|_x . \tag{1}$$

In particular, McCaffrey [12,44], Heskestad [17,18] and Alpert & Ward [45] reported empirical correlation models for both flame heights, given by the relations

$$h_{cf}|_x = 0.2\dot{Q}^{\frac{2}{5}}\Big|_{McCaffrey} \; ; \; 0.235\dot{Q}^{\frac{2}{5}} - 1.02 D_f \Big|_{Heskestad} \; ; \; 0.174 k \dot{Q}^{\frac{2}{5}}\Big|_{A\&W}, \tag{2}$$

$$h_{if}|_x = 0.08\dot{Q}^{\frac{2}{5}}\Big|_{McCaffrey} \; ; \; 0.07\dot{Q}^{\frac{2}{5}}\Big|_{Heskestad} \; ; \; 0.07\dot{Q}^{\frac{2}{5}}\Big|_{A\&W}, \tag{3}$$

which are all based on the computation of the heat release rate, $\dot{Q}$. Notice that in the continuous flame height correlation proposed by Heskestad, a function of the flame/nozzle equivalent diameter, $D_f$, is included, while the Alpert & Ward correlation takes into account the fire confinement constant, $k = 1,2,3$ [45].

Alternatively, Delichatsios [23] developed a correlation by considering the buoyancy effects. In terms of the flame Froude number, $Fr_f$, and the dimensionless flame length, $L^*$, the luminous flame height, $h_{Lf}|_{Fr}$, can be determined from the relation

$$h_{Lf}|_{Fr} = \frac{L^* d_j (\rho_e/\rho_\infty)^{\frac{1}{2}}}{f_s}, \tag{4}$$

where the dimensionless flame length is a function of the flame Froude number, which in turn is determined through several other parameters, such as the fuel jet diameter, $D_{fj}$, the fuel and ambient density, $\rho_e$ and $\rho_\infty$, respectively, the fuel velocity, $v_e$, the stoichiometric mixture fraction, $f_s$, the temperature gradient between the flame and its surroundings, $T_f$ and $T_\infty$, respectively, and the gravitational constant acceleration, $g$, as given by following expressions

$$L^* = \frac{13.5 Fr_f^{\frac{2}{5}}}{\left(1 + 0.07 Fr_f^2\right)^{\frac{1}{5}}}, \tag{5}$$

$$Fr_f = \frac{v_e f_s^{\frac{3}{2}}}{\left(\frac{\rho_e}{\rho_\infty}\right)^{\frac{1}{4}} \left[\frac{\Delta T_f}{T_\infty} g D_{fj}\right]^{\frac{1}{2}}}. \tag{6}$$

At small Froude numbers ($Fr_f < 5$), Eq. (5) simplifies to $L^* = 13.5 Fr_f^{2/5}$, with a clear tendency to approach the buoyancy-dominated limit. Conversely, when $Fr_f > 5$, the dimensionless flame length predicted by this correlation asymptotically approaches the momentum-dominated dimensionless value, given by $L^* = 23$ [23].

### B. Flame Temperature

In combustion processes, the heat released during the exothermic reaction raises the temperature of the combustion products, reaching its maximum value at a certain region of the flame. When the reaction is carried out under the condition that the combustion process is complete, there is not heat loss to the surroundings (i.e., $\delta Q = 0$). This temperature is known as the adiabatic flame temperature, $T_{adiab}$. Since there is nothing that confines the gas, it is further assumed that the substances burn at a constant pressure. The energy transfer of the reacting system is determined using the first law of thermodynamics:

$$\delta Q = 0 \rightarrow dH = 0 \rightarrow H^b = H^u, \tag{7}$$

where the burnt and unburnt gases are denoted by superscripts *b* and *u*, respectively, *Q* is the heat, and *H* are the enthalpies. Furthermore, the molar enthalpies of the burnt and unburnt gases often differ, because the amount of molecules usually changes in a chemical reaction. Thus,

$$h^u = \sum_{j=1}^S w_j^u h_j^u = \sum_{j=1}^S w_j^b h_j^b = h^b, \tag{8}$$

where *h* denotes enthalpy, *j* the species, and *w* is the work. For constant pressure the following relation holds:

$$h_j^b = h_j^u + \int_{T_u}^{T_b} c_{p,j} dT. \tag{9}$$

It is stated that the sum of the enthalpies of all reactive species at inlet temperature, $T_{ref}$, must equal the sum of the enthalpies of all product species at a given final temperature. Integrating over the temperature, the above expression becomes:

$$\sum H_P(T) = \sum n_j \left[\Delta H_{f,j}^0 + \overline{Cp_j}(T_{adiab} - T_{ref})\right]. \tag{10}$$

Consequently, the adiabatic flame temperature can be determined through an iteration process employing the standard enthalpy of formation, $\Delta H_{f,j}^0$, the mean heat capacity data available, $\overline{Cp_j}$, and the molar concentration, $n_j$, of the chemical species. Equation (10) quantifies the maximum temperature achieved by the products and is determined from the energy balance of the reaction at equilibrium. In addition, this temperature depends only on the fuel-oxidizer ratio, since both the formation enthalpy and the specific heat are functions of the involved species.

In summary, the analysis presented in this section provides qualitative estimations of the magnitudes that the flame height and temperature are expected to reach.

## III. METHODOLOGY

### A. Experimental Design and Setup

To validate the reliability of the present methodology, some preliminary steps were implemented to reduce the number of tests, while achieving a less restricted validity field. These steps include: (i) determining the experimental units and factors, (ii) defining the experimental procedure and conducting a pilot test and (iii) specifying the statistical design.

Based on the previous analysis, the salient experimental factors are determined to be: (i) the fuel-oxidizer mixture, (ii) the volumetric fuel flow and (iii) the visualization plane, with their corresponding levels as listed in Table I.

Table I: Primary factors and corresponding levels of the experimental design.

| Factors | Levels |
| --- | --- |
| Fuel-Oxidizer | LP Gas-Air |
| Volumetric Flow [cc/min] | 350, 650, 950, 1200, 1500, 1800 |
| Visualization Planes | Lateral, Frontal, Angled |

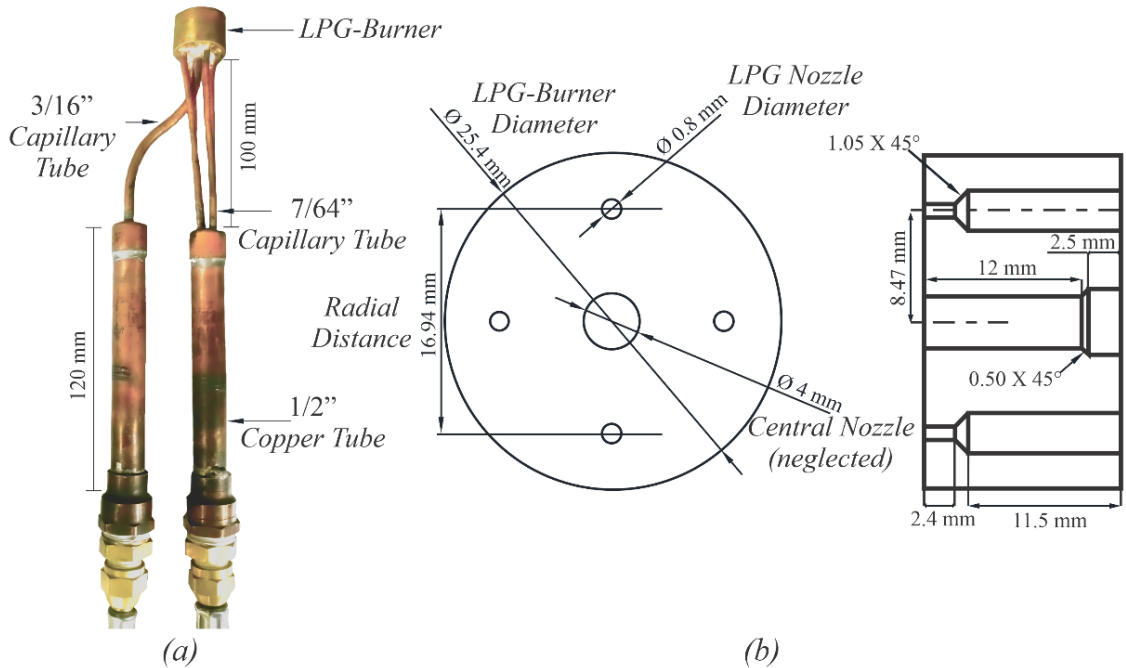

Fig. 2: Central-peripheral fuel injection system: (a) distribution setup and (b) gas-burner configuration.

The experimental rig is shown in Fig. 2. It utilizes a gas-burner, central-peripheral fuel injection system to study the behaviour and define the structure of a non-confined diffusion flame (Fig. 2a). The fuel injection system for this particular experimental setup was built based on a blowtorch distribution design. The proposed setup employs as reacting species LPG (as fuel) and air (as the oxidizing agent). For this experimental study, the LPG is composed approximately of 60% propane and 40% butane (see Table II). This composition has thermo-chemical and reaction properties very similar to those reported by Mishra and Rahman [46]. The experimental evaluation was designed to determine the flame height for six different fuel flows: (a) 350 cc/min, (b) 650 cc/min, (c) 950 cc/min, (d) 1200 cc/min, (e) 1500 cc/min and (f) 1800 cc/min as listed in Table I. The fuel injection system makes use of a four-nozzle port array instead of the whole 4-Lug-Bolt setup [1]. This arrangement consists of four 0.8 mm peripheral nozzles in a 4 x 16.94 mm circularly arranged configuration for a 25.4 mm diameter gas-burner, as depicted in Fig. 2b. The 4 mm central nozzle was not employed here and was left aside to complement an ongoing research project involving the Lug-Bolt configuration. The proposed diameters of the gas-burner were calculated in order to maintain a stoichiometric mixture fraction $f_s$=0.06047 between the fuel and the oxidizing agent.

Table II: Properties of LPG-Air at 293.15 K and 0.7647 atm.

| Case Study | LPG Volumetric Flow [cc/min] | LPG Injection Velocity [m/s] | LPG Mass Flow [kg/s] | Injection Re GLP | Stoichiometric Air Mass Flow needed [kg/s] | Air–Fuel Ratio, $(A/F)_{stoich}$ |
|---|---|---|---|---|---|---|
| a | 350 | 2.901 | 9.06x10$^{-6}$ | 498.1 | 1.70x10$^{-4}$ | |
| b | 650 | 5.388 | 1.68x10$^{-5}$ | 925 | 3.16x10$^{-4}$ | |
| c | 950 | 7.874 | 2.45x10$^{-5}$ | 1351.6 | 4.62x10$^{-4}$ | 15.53 |
| d | 1200 | 9.947 | 3.10x10$^{-5}$ | 1707.4 | 5.83x10$^{-4}$ | |
| e | 1500 | 12.434 | 3.88x10$^{-5}$ | 2134 | 7.29x10$^{-4}$ | |
| f | 1800 | 14.921 | 4.65x10$^{-5}$ | 2561 | 8.75x10$^{-4}$ | |

In combustion theory, the stoichiometric relation for a hydrocarbon fuel represented by $C_xH_y$, is given by

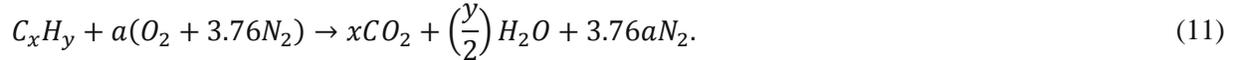

$$C_xH_y + a(O_2 + 3.76N_2) \rightarrow xCO_2 + \left(\frac{y}{2}\right)H_2O + 3.76aN_2. \qquad (11)$$

For simplicity, in the above reaction the air has a composition of 21%$O_2$ and 79%$N_2$. It is assumed that this reaction is balanced where a=x+(y/4). Then, the stoichiometric air/fuel ratio is $(A/F)_{stoich}=(m_{ox}/m_{fuel})=4.76a(MW_{air}/MW_{fuel})$, where $MW_{air}$ and $MW_{fuel}$ are the molecular weights of air and fuel, respectively. The energy production of LPG is established by the overall reaction:

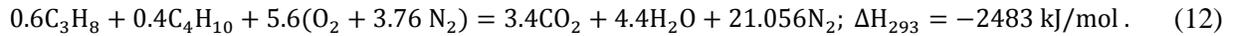

$$0.6C_3H_8 + 0.4C_4H_{10} + 5.6(O_2 + 3.76\ N_2) = 3.4CO_2 + 4.4H_2O + 21.056N_2;\ \Delta H_{293} = -2483\ kJ/mol. \qquad (12)$$

The instrumentation employed for this experimental setup is displayed in Figs. 3 and 4. The fuel supply system (Fig. 3a) comprises a Parker pressure regulation valve Mod. N400S for robust control of the volumetric flow and a Dwyer flowmeter Mod. RMA 14 (4% F.S.) employed for fine adjustment and measuring.

To measure the flame height, an array of three high-definition cameras were installed at a 1 m radius from the gas-burner to take frontal (0°), angled (45°) and lateral (90°) images of the flame, as seen in Fig. 3b. This arrangement was fitted with a rig of interconnected linear electric actuators placed on the top of each camera, which simultaneously trigger the shutter. The complete diagram of the proposed experimental setup is shown in Fig. 4. In addition, a Fluke-Ti55FT thermal-imaging camera was placed in the workbench frontal-plane at 1 m from the gas-burner to avoid emissivity errors [47–49]. To complement this, the temperature of the flame was also measured employing a Heraeus pyrometer Mod. DT-400 (1% F.S.) fitted with a Type C alloy thermocouple probe of Tungsten-Rhenium, positioned at the centreline above the nozzle of the observed flame. These instruments and their setup are shown in Fig. 3b and Fig. 4, respectively. After conducting a pilot test, the preliminary measure output obtained is compared against theoretical flame height estimations, as initially expected values.

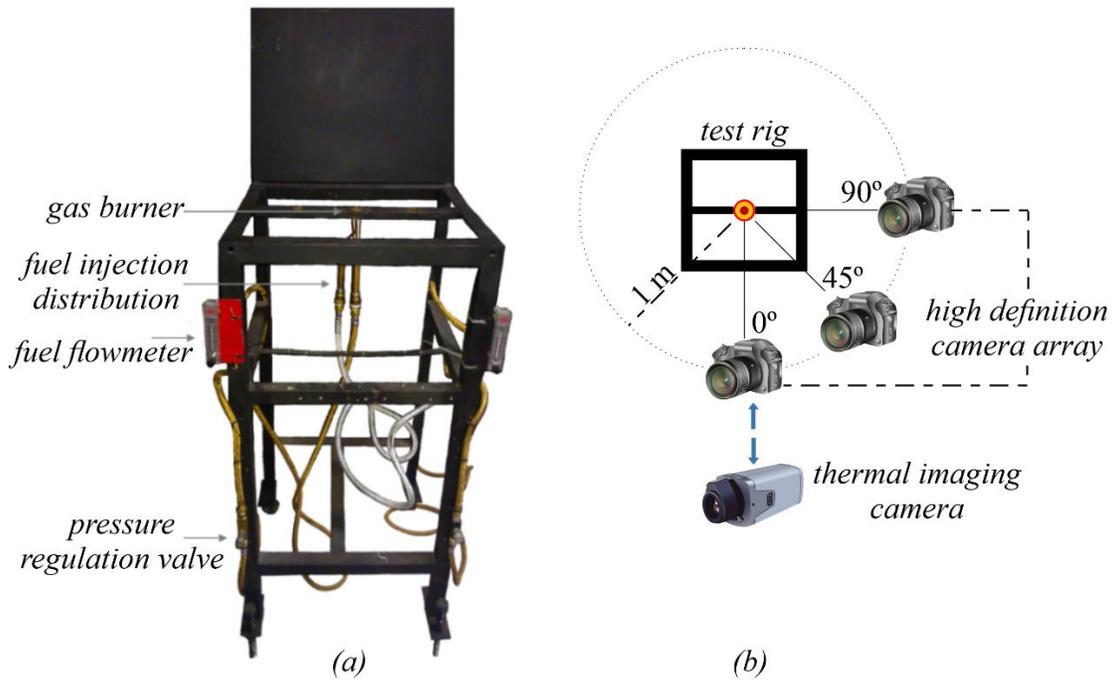

Fig. 3: Experimental setup; (a) test rig and (b) camera-visualization array.

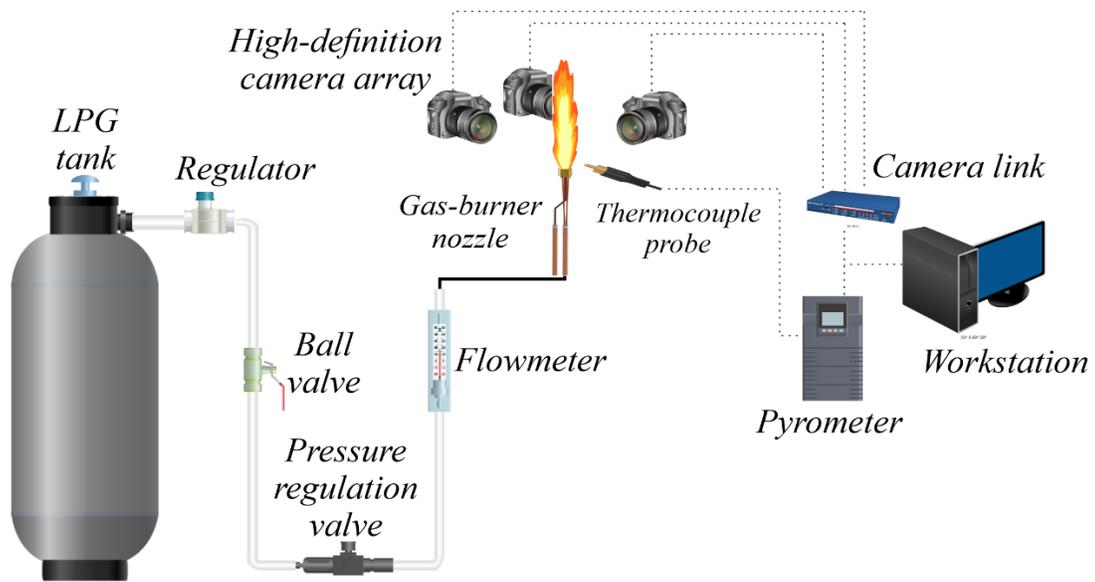

Fig. 4: Diagram of the experimental setup.

The difference between the data sets revealed the existence of additional nuisance factors, particularly, a light air current inside the chamber where the experiment was performed and sunlight coming indirectly from a window. Several measures were taken in order to suppress these perturbing agents, such as reducing and isolating the area of the chamber and employing thick black curtains. Flame stabilization was obtained during a 10-15 s period after ignition to ensure proper image capture. Thereafter, a new test was performed and the output was compared once more against the benchmark with no significant differences.

These additional nuisance factors are deemed to be noise as they possess unknown magnitudes. Therefore, they are uncontrollable and of no interest to the research topic. Nonetheless, a randomized complete block design (RCBD) is implemented to reduce their contribution to the experimental errors, thus leading to greater accuracy. In addition, the run sequence of the treatments is determined randomly to achieve a higher significance in terms of both forward-carry and instrumentation error.

**B.    Convolution-Based Image Processing**

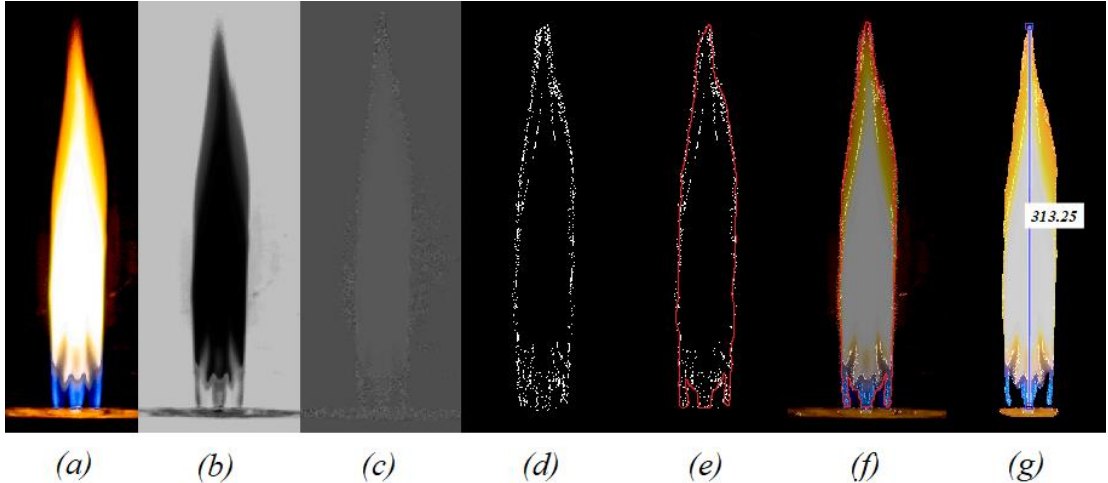

Fig. 5: Image processing algorithm output.
(a) Augmented image, (b) first convolution, (c) second convolution, (d) edge detection, (e) edge outline, (f) image superposition/blending and perimeter smoothing and (g) length measurement/output image.

For this specific experiment, an image-processing algorithm was developed and programmed, borrowing concepts and fundamentals from different implementations. This was divided into four main layers: (i) image augmentation, to enhance the features of the image and facilitate landmark detection, (ii) convolution, to assign importance to the objects in the image and differentiate one from the other based on the pixel value, (iii) edge detection, to reduce the image into a form that makes it easier to process without losing relevant features, and (iv) digital mapping, for image superposition as well as a projection into an *xy*-axes coordinate map, to calculate the difference between the largest and smallest y-coordinates. The corresponding relevant outputs are shown in Fig. 5. For image augmentation, the requirements were minimal and only contrast increase was applied to the input image to enhance shape detection (Fig. 5a).

Subsequently, the resulting upgraded image is subjected to the convolution operation to extract the high-level features, mainly the edges of the input image through the utilization of a user-defined $n \times n$ kernel matrix, $\omega$, and corresponding filters, $K_n$. The employed kernel and filters are given by

$$\omega = \begin{bmatrix} -1 & -1 & -1 \\ -1 & 8 & -1 \\ -1 & -1 & -1 \end{bmatrix} ; K_1 = 3 \rightarrow \begin{bmatrix} R \\ G \\ B \end{bmatrix} ; K_{2 \rightarrow n} = 1 . \qquad (13)$$

This process is made for every channel of the input image. The resulting values of each filter are added and the product is arranged in an output activation map, thus taking the 3-channel image input and converting it into a one-depth convoluted feature output, as shown in Figs. 5b and 5c.

The kernel is slid across the input image based on a defined stride, $S$, and a zero-padding value, $P$. As a first approach and for validation purposes, a stride $S = 1$ is employed and a full zero-padding is applied. As shown below in Eq. (14), the convolution operation is performed between the weights of the kernel. The pixel values corresponding to the image section are evaluated and the product is then assigned to an output image representing the new pixel values, resulting in an enhanced feature (Figs. 5b and 5c).

$$\begin{bmatrix} x_{11} & x_{12} & \cdots & x_{1n} \\ x_{21} & x_{22} & \cdots & x_{2n} \\ \vdots & \vdots & \ddots & \vdots \\ x_{m1} & x_{m2} & \cdots & x_{mn} \end{bmatrix} * \begin{bmatrix} y_{11} & y_{12} & \cdots & y_{1n} \\ y_{21} & y_{22} & \cdots & y_{2n} \\ \vdots & \vdots & \ddots & \vdots \\ y_{m1} & y_{m2} & \cdots & y_{mn} \end{bmatrix} = \sum_{i=0}^{m-1} \sum_{j=0}^{n-1} x_{(m-1)(n-j)} y_{(1+i)(1+j)} \qquad (14)$$

In a similar way, edge detection is performed over the down-sampled convoluted feature to extract relevant image elements as functions of both the magnitude and direction of the gradient as, for example, the strength of the edges. For this experiment, the well-known Sobel operator [50–52] was employed with the vertical edge direction being favoured to give the resulting shape of the observable flame.

A convolution operation is made between the kernel and the input image, and the edge strength is obtained as depicted in Fig. 5d. Subsequently, a threshold value is applied to decide whether edges are present or not at an image point. A relatively high threshold was employed to reduce the noise as well as the possible detection and inclusion of irrelevant features in the output image.

Once the binary edge gradient image is obtained an outline is drawn around the flame structure, as shown in Fig. 5e. Subsequently, the original image is fused with the drawn flame contour through image superposition and such an outline is then smoothed through morphological structuring of the blended image to further increase the adjustment of the created edge outline (see Fig. 5f).

The image is then superimposed on a Cartesian plane as a function of the pixel position. The highest y-coordinate of the generated contour is subtracted from the lowest one in order to determine the flame height in pixel length. For display purposes, the algorithm shows the original image with everything outside the edge contour being cropped and the length measurement overlaid, as shown by Fig. 5g.

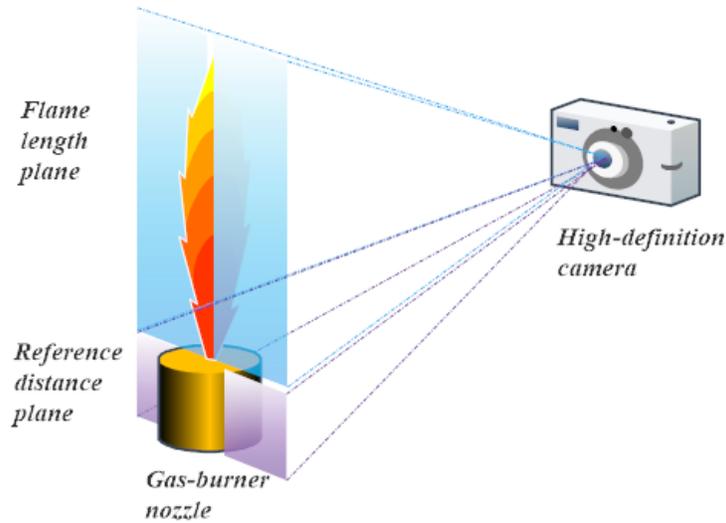

Fig. 6: Reference length and flame length plane.

The actual flame length is measured indirectly by employing a known reference distance and its pixel length equivalent, using the expression

$$h_{Lf}\big|_{exp} = \frac{(h_{Lf,pixels} L_{Nozzle,real})}{L_{Nozzle,pixels}}, \qquad (15)$$

where the 1-inch diameter gas-burner nozzle located in the same plane as the actual flame is used to avoid parallax-induced errors (see Fig. 6).

### C.    Post-Processing of the Experimental Data

To properly validate the measurements obtained from the proposed methodology, a post-processing stage is implemented here, based on a thorough statistical analysis of the effect of the factors, of the observations and their corresponding errors, the uncertainty analysis and finally a comparison between the theoretical values and the actual experimental observations.

#### 1.    *Effects of the experimental factors and distribution fitting*

In order to assess the effects of the factors on the experimental design, an analysis of variance (ANOVA) is performed on the response dataset [53,54]. The test statistics $F$ is computed in order to quantify the ratio of the column variance to the row variance as well as their interaction. Based on this statistics, its probability value, $P_{(F)}$, is then determined through the $F$-distribution to differentiate the results obtained from chance due to the sampling from those that are statistically significant, in terms of the proposed null hypothesis significance testing ($NHST$), namely

$$NHST \rightarrow \begin{cases} H_O: \theta - \theta_O = 0 \\ H_a: \theta - \theta_O \neq 0 \end{cases}. \tag{16}$$

The previously assigned significance level, $\alpha$, and the probability value are compared and, based on the outcome, it is possible to either fail to reject the null hypothesis, $H_O$, given $P_{(F)} > \alpha$, or reject $H_O$, thus accepting the alternate, $H_a$, based on $P_{(F)} < \alpha$.

However, both the $F$ statistics and its probability value must be used in combination in order to properly evaluate and decide the significance of the results since the probability value provides a broader insight of the variability as a set, and the $F$ statistics shows the variability due to underlying conditions attributed to the sources [53,54]. Furthermore, to confirm that the output dataset exhibits a behaviour similar to normality, both the Shapiro-Wilk [55] and the Kolmogorov-Smirnov [56–58] distribution fitting tests are employed.

#### 2.    *Statistical tolerance of the experiment*

Once the effects of the factors have been determined and the observations have been confirmed to come from a normal distribution, the quality of the results, which are associated to the standard error of the measurement (SE) will be examined for each observation dataset. Using as a metric the root sum of squares, $RSS$, defined as

$$RSS = \sqrt{\sum_{i=1}^{R} \sum_{j=1}^{C} (SE_{ij})^2}, \tag{17}$$

it will be possible to obtain the variation of each factor and combine them, thereby allowing to determine the probability that the analysed unit falls within acceptable limits, i.e., within the statistical tolerance.

#### 3.    *Uncertainty measurement*

Given its sample size and number of replications and considering that the data were self-acquired and exhibit a normal behaviour, it is assumed that the systematic error was transformed into a random error.

Therefore, a Type A uncertainty analysis must be performed. For this study, where multiple repeatability tests were conducted, the method of pooled variance, $\sigma_p^2$, is employed to combine the standard deviations, $\sigma_p$. Then, from the replications of each treatment, $n$, the standard uncertainty, $u(y_i)$, is determined using the expression

$$u(y_i) = \frac{\sigma_p}{\sqrt{n}}. \tag{18}$$

Furthermore, to maintain the measurement of uncertainty within the 95% CI (confidence interval), a coverage factor $k \approx 2$ is assigned to obtain an expanded uncertainty, $U$, given by [59,60]

$$U = k(u(y_i)). \tag{19}$$

### 4. Measurement error and adjustment

After validation of the experimental design and output has been obtained, the results will be compared to the forecasts obtained through available benchmarks [12,17,18,44,45]. This analysis is conducted to determine the bias of the resulting data set in terms of the physical phenomena.

To assess the quality of the information obtained, the bias, $MAE$, and the associated errors, $RMSE$ and $MAPE$, are determined by the following relations

$$MAE = \left(\frac{1}{n_c}\right) \sum |y_f - y_o|, \tag{20}$$

$$RMSE = \sqrt{\left(\frac{1}{n_c}\right) \sum (y_f - y_o)^2}, \tag{21}$$

$$MAPE = \left(\frac{1}{n_c}\right) \sum \left|\frac{y_f - y_o}{y_o}\right| 100, \tag{22}$$

based on the observations, $y_o$, and the forecasts, $y_f$, where $n_c$ is the sample size for each treatment and the bias, $MAE$, establishes the direction of the error of the estimates, thus identifying any particular underlying conditions attributed to any issue in the system or procedure used to obtain such estimates. The root mean square error, $RMSE$, determines the overall distance between the observed data set and the forecasts, while the mean absolute percentage error, $MAPE$, provides the ratio of accuracy in terms of the expected error when designing the experiment.

## IV. RESULTS AND DISCUSSION

### A. Luminous flame height measurement

From the experimental procedure outlined in the previous section, 576 images of the luminous flame were obtained. The proposed algorithm was then used to process these images and the overall results obtained are condensed in Tables III and IV, while the resulting images are displayed in Fig. 7.

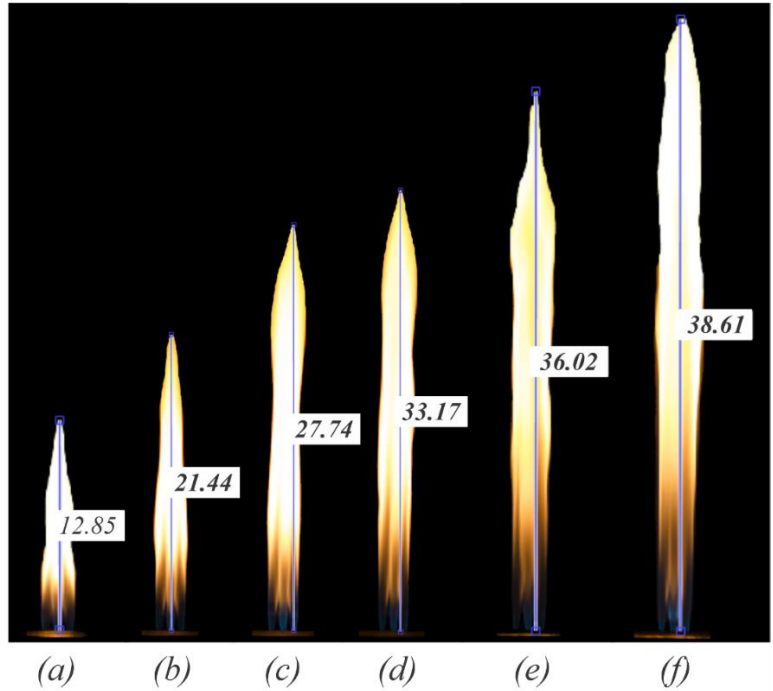

Fig.7: Output measurements from the high-definition camera.
(a) 350 cc/min, (b) 650 cc/min, (c) 950 cc/min, (d) 1200 cc/min, (e) 1500 cc/min and (f) 1800 cc/min.

Table III lists the main statistics of the obtained flame height measurements, where the mean flame height can be determined for every volumetric fuel flow analysed. The obtained overall mean height values are: (i) 12.85 cm for 350 cc/min, (ii) 21.44 cm for 650 cc/min, (iii) 27.74 cm for 950 cc/min, (iv) 33.17 cm for 1200 cc/min, (v) 36.02 cm for 1500 cc/min and (vi) 38.61 cm for 1800 cc/min.

Table III: Main results/statistics of the luminous flame height measurements.

| Camera direction | Volumetric fuel flow | Replications | Mean flame height | Standard deviation | Variance | Range | Coeff. Of variation |
|---|---|---|---|---|---|---|---|
| Lateral | 350 | 32 | 12.389 | 1.456 | 2.120 | 4.931 | 11.94 |
| | 650 | 32 | 20.209 | 2.213 | 4.897 | 9.191 | 11.12 |
| | 950 | 32 | 26.409 | 1.653 | 2.734 | 9.457 | 6.36 |
| | 1200 | 32 | 31.988 | 2.260 | 5.108 | 10.213 | 7.17 |
| | 1500 | 32 | 35.960 | 1.661 | 2.760 | 6.882 | 4.69 |
| | 1800 | 32 | 38.705 | 2.059 | 4.241 | 7.634 | 5.40 |
| Angled | 350 | 32 | 13.466 | 1.075 | 1.156 | 4.679 | 8.11 |
| | 650 | 32 | 22.445 | 1.639 | 2.688 | 5.599 | 7.42 |
| | 950 | 32 | 28.924 | 1.458 | 2.126 | 7.362 | 5.12 |
| | 1200 | 32 | 34.302 | 1.457 | 2.123 | 5.129 | 4.31 |
| | 1500 | 32 | 36.864 | 1.574 | 2.478 | 5.844 | 4.33 |
| | 1800 | 32 | 38.545 | 1.811 | 3.282 | 6.951 | 4.77 |
| Frontal | 350 | 32 | 12.700 | 0.835 | 0.697 | 3.314 | 6.68 |
| | 650 | 32 | 21.680 | 1.525 | 2.327 | 7.053 | 7.14 |
| | 950 | 32 | 27.899 | 1.638 | 2.685 | 6.288 | 5.96 |
| | 1200 | 32 | 33.247 | 1.707 | 2.916 | 9.215 | 5.21 |
| | 1500 | 32 | 35.244 | 2.037 | 4.150 | 9.811 | 5.87 |
| | 1800 | 32 | 38.600 | 1.807 | 3.267 | 6.542 | 4.75 |

Furthermore, Table III contains the range and the coefficient of variation, $CC_V$. The former shows that for the upper and lower limits of the volumetric fuel flow, the interval remains relatively similar. However, when observing the behaviour for intermediate fuel flows, subtle differences arise, which lead to infer that the lateral camera presents the highest average range with 8.051 cm, followed by the frontal camera with 7.037 cm and the angled camera with 5.927 cm. Meanwhile, the coefficient of variation shows that for the 350 cc/min case at low fuel flows, the variability of the data is the highest for every camera. Nonetheless, this variability diminishes as the fuel flow increases, reaching ~ 50% reduction when measuring the upper bound limit and entailing a stabilization of the luminous flame front. Furthermore, it is seen that the lateral camera presents a slightly higher average $CC_V$ of 7.78%, followed by the frontal camera with 5.93% and the angled camera with 5.67%. This implies that the instabilities are either more visible from this plane or that the camera employed was susceptible to the variations on the perceivable flame height.

In addition, by calculating the estimated differences between each pair of means: (i) angled-lateral with 1.481 cm, (ii) angled-frontal with 0.862 cm and (iii) frontal-lateral with 0.618 cm, it was found that even though there are slight differences between the cameras, given that the frontal-lateral one is the lowest, the 45° plane measurements stand out from the others. This is because the angled camera was able to capture uneven flame structure features, like flame skewness, tip flickering and/or flame kernels still attached to the flame front, due to the blending of the remaining planes.

Through an analysis of variances, Table IV decomposes the variability of the obtained experimental flame height measurements into the contributions of the factors. This measurement is performed after having removed the effect of all other factors, and their statistical significance is tested.

Table IV: Two-way ANOVA of the resulting experimental values.

| Source | Sum of squares | Degrees of freedom | Mean square | F-Ratio | P-Value |
|---|---|---|---|---|---|
| *Main effects* | | | | | |
| R:Camera Direction | 212.486 | 2 | 106.243 | 35.790 | ~ 0.000 |
| C:Volumetric Flow | 45679.876 | 5 | 9135.975 | 3077.660 | ~ 0.000 |
| *Interactions* | | | | | |
| I:RC | 120.669 | 10 | 12.066 | 4.070 | ~ 0.000 |
| *Residual* | | | | | |
| Error | 1656.410 | 558 | 2.968 | | |
| Total | 47669.442 | 575 | | | |

As shown in Table IV all the resulting *P*-Values are below the previously selected significance level $\alpha = 0.05$, which states that within the 95% CI all the factors as well as their interaction have a statistically significant effect on the experimental unit.

In spite of that, on analysing the *F* statistics, the volumetric fuel flow is the factor with the highest contribution to the output variability, since the larger the fuel load, the more intense the flame becomes. In contrast, a relatively low *F* statistic is obtained for the camera direction, which implies that even though the position of the camera exerts an effect on the output, the variations of the flame height between the planes are minimal compared to the fuel flow. However, these slight variations are attributed to the inability of the camera to distinguish between the constant chemical and the intermittent luminescent reactions of the combustion phenomena. In addition, the variation due to the interaction is practically negligible, implying that even though differences in the fuel flow are more perceivable from a certain plane, these two factors are not particularly correlated.

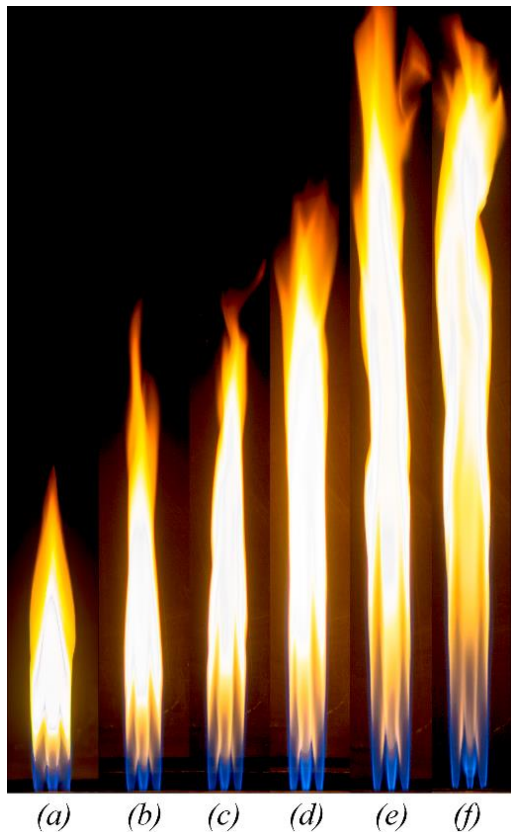

Fig. 8: Overall average flame structure as a function of the fuel flow.
(a) 350 cc/min, (b) 650 cc/min, (c) 950 cc/min, (d) 1200 cc/min, (e) 1500 cc/min and (f) 1800 cc/min.

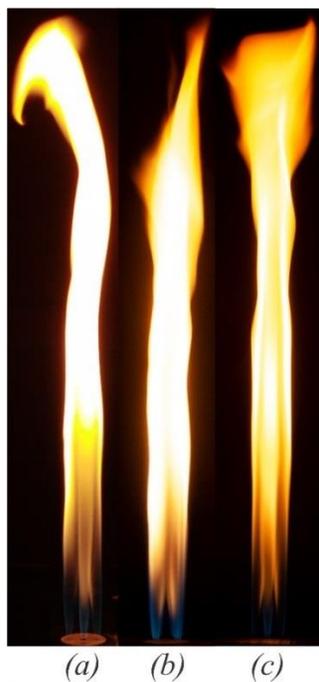

Fig.9: Overall average flame structure as a function of the camera direction.
(a) lateral, (b) angled and (c) frontal

Figures 8 and 9 show that the luminosity in the lower half of the flame is more stable, while in the upper half instabilities, such as flame crookedness, flickering and intermittent luminescence, can be observed as the result of the upward shedding of vortex structures located on the flame boundaries [2,42].

Once the factors of the experiment and their effects are accounted for, the output values must be evaluated in order to determine the effectiveness of the proposed experimental design. This is achieved by performing a distribution fitting of the observations. The calculated statistical tolerance and uncertainty evaluation are given in Tables V, VI, and VII, respectively.

The results of the distribution fitting are listed in Table V. Since the $P$-Values, $P_{(SW)}$ and $P_{(KS)}$, among the tests performed are greater than $\alpha = 0.05$, the possibility that these observations come from a normal distribution cannot be rejected within a 95% CI.

After having determined that the observations come from a normal distribution, the statistical tolerance can be computed. This calculation yields a $RSS_{SE} = 1.2$ cm (see Table VI), which is lower than the 3 cm limit assigned during the experimental design. This value indicates the overall error that can be present during the treatments. The upper and lower error bounds are 9.89% and 3.28%, respectively.

Table V: Test for normality of the output values.

| Camera direction | Volumetric fuel flow | Shapiro-Wilk P-Value | Kolmogorov-Smirnov P-Value |
|---|---|---|---|
| Lateral | 350 | 0.4210 | 0.6231 |
| | 650 | 0.3888 | 0.7956 |
| | 950 | 0.1364 | 0.5239 |
| | 1200 | 0.7483 | 0.9824 |
| | 1500 | 0.6427 | 0.8562 |
| | 1800 | 0.3813 | 0.6666 |
| Angled | 350 | 0.8265 | 0.9360 |
| | 650 | 0.3116 | 0.9539 |
| | 950 | 0.1335 | 0.7060 |
| | 1200 | 0.3089 | 0.9505 |
| | 1500 | 0.5521 | 0.9318 |
| | 1800 | 0.5228 | 0.8600 |
| Frontal | 350 | 0.5237 | 0.6974 |
| | 650 | 0.5978 | 0.9386 |
| | 950 | 0.3719 | 0.8693 |
| | 1200 | 0.1557 | 0.6841 |
| | 1500 | 0.6168 | 0.8034 |
| | 1800 | 0.3563 | 0.9117 |

Table VI: Standard error of the measurement and statistical tolerance.

| Camera direction | Volumetric fuel flow | Standard error | Root sum of squares |
|---|---|---|---|
| Lateral | 350 | 0.257 | |
| | 650 | 0.391 | |
| | 950 | 0.292 | |
| | 1200 | 0.399 | |
| | 1500 | 0.293 | |
| | 1800 | 0.364 | 1.271 |
| Angled | 350 | 0.190 | |
| | 650 | 0.289 | |
| | 950 | 0.257 | |
| | 1200 | 0.257 | |
| | 1500 | 0.278 | |
| | 1800 | 0.320 | |

|  | 350 | 0.147 |
|---|---|---|
|  | 650 | 0.269 |
|  | 950 | 0.289 |
| Frontal | 1200 | 0.301 |
|  | 1500 | 0.360 |
|  | 1800 | 0.319 |

The analysis also gives the statistical tolerance for each camera, with a $RSS_{SE}$ of 0.82 cm for the lateral, 0.70 cm for the frontal and 0.65 cm for the angled camera. In general, these findings reveal that the intrinsic error associated to each camera is relatively low. The difference between the largest and smallest $RSS_{SE}$ is just 0.17 cm, which suggests that most of the variability should be attributed solely to the flame skewness, the flame flickering and the occasional capture of the kernels still attached to the flame front.

On the other hand, the error due to changes of the volumetric fuel flow were computed, yielding $RSS_{SE}$ values of 0.35 cm, 0.55 cm, 0.48 cm, 0.56 cm, 0.54 cm and 0.58 cm for the 350, 650, 950, 1200, 1500 and 1800 cc/min regimes, respectively. These variations are attributed to the inability of the cameras to properly capture the flickering within the flame structure as produced by the upward shedding vortices. However, since they all lie within the initial limit proposed of ~2cm, and given that the maximum variation between them is 0.23 cm, the output is considered to be acceptable.

The results of the Type A uncertainty measurement are shown in Table VII. The pooled standard deviation, as obtained from the pooled variance method, and the standard uncertainty are $\sigma_P = 1.695$ cm and $u(y_i) = 0.299$ cm, respectively. To maintain such an uncertainty measurement within the 95% CI, a coverage factor is assigned to give $U = \pm 0.599$ cm, which can be interpreted as an uncertainty measurement of $\pm 11.33\%$, a value well within the 20% limit assigned during the experimental design.

Since $2U \approx RSS_{SE}$, both the experimental design and the procedure presented in this paper can be considered to be acceptable, based on the fairly balanced error-uncertainty ratio that was obtained. This implies that the observations tend to remain relatively centred and the obtained measurement of uncertainty does not favour a particular direction, thereby suggesting that the systematic error effectively became a random error. This clarifies why the persistent bias is related to the random behaviour of the combustion process.

Table VII: Type A uncertainty measurement results.

| Pooled variance | Pooled standard deviation | Standard uncertainty | Coverage factor | Expanded uncertainty | Relative uncertainty |
|---|---|---|---|---|---|
| 2.875 | 1.695 | 0.299 | 2 | 0.599 | 11.33% |

Finally, in order to validate the results obtained using the present methodology, the experimental observations are compared to theoretical forecasts obtained from the literature and then evaluated to determine how well they fit the behaviour described by the empirical correlations.

As shown in Fig. 10, the observed experimental data fit the current theoretical models for flame height measurement. In particular, the agreement is good for the largest fuel flow regimes, i.e. 950, 1200, 1500 and 1800 cc/min, where the maximum relative error amounts to 10% for the 950 cc/min fuel flow. Alternatively, for the remaining fuel flows the maximum relative error rises to 23% and 60%, for the 650 cc/min and 350 cc/min cases, respectively. This is mainly because diffusion flames at low Reynolds numbers tend to become unstable under the effect of buoyancy-induced vortices, leading to bulk flickering and thus making it more difficult to measure.

Furthermore, as stated in Section II, the available correlations were developed and tested for particular combustion setups, specifically pool fire and co-flow. However, for lug-bolt/port-array configurations, the diffusion mechanism is rather different, since (i) the fuel diffuses in between streams, developing individual potential cores, hence the absence of a bulk flame potential core, (ii) these streams, as seen in the figures, remain relatively undisturbed until the flame streams join in the mixing zone, conforming the flame front, and (iii) at every fuel stream the fuel diffuses outward, while the oxidizer diffuses on the flame surface. This induces the stream mixing zone to move inwards and towards the centre, resulting in inter-diffusion along the vertical direction.

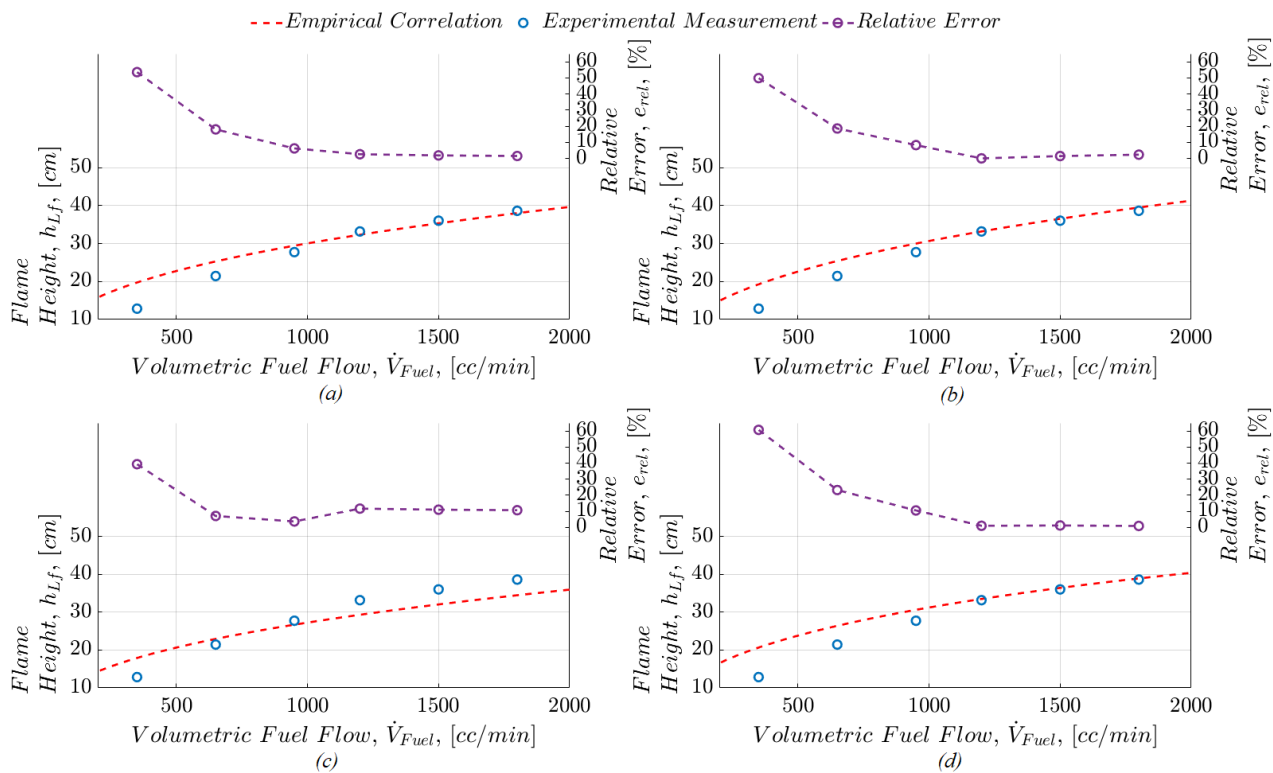

Fig. 10: Relative error and adjustment of the mean observed data.
(a) McCaffrey, (b) Heskestad, (c) Alpert & Ward and (d) Froude number.

These combined phenomena lead to a fuel-oxidizer diffusion downstream in such a mixing process, where the flame is ignited, resulting in a lengthwise lower interface of the mixing zone. Therefore, a difference between observations and forecasts was already expected and the experiment was designed accordingly. However, the magnitude of the errors, particularly for the lower fuel flows, reveals the need to develop a correlation that fits this specific setup, proving that the combustion dynamics is greatly affected by the gas-burner configuration.

This idea is further corroborated by the scatter plots (Fig. 10), where it is seen that for these fuel flow regimes the observation remains below the predicted value, with an average of 6.5 degrees for the 350 cc/min flow and 3.5 degrees for the 650 cc/min. The persisting bias leads to observations consistently lower than the predictions, which may be attributed to the aforementioned particularities of the employed gas-burner configuration.

Table VIII: Adjustment of the observed data against the current theoretical modelling.

| Source | Mean absolute error | Root mean square error | Mean absolute percentage error |
|---|---|---|---|
| McCaffrey | 1.725 | 3.337 | 13.962 |
| Heskestad | 2.353 | 3.247 | 13.420 |
| Alpert & Ward | 1.063 | 3.566 | 13.881 |
| Froude number | 2.790 | 3.968 | 16.221 |

As shown in Table VIII, the overall adjustment of the models remains fairly appropriate. The mean absolute error shows that the Froude number-based correlation presents the highest average differences between forecasts and observations with 2.79 degrees, followed by the Heskestad correlation with 2.35 degrees, the McCaffrey correlation with 1.72 and the Alpert & Ward correlation with 1.06 degrees. However, when analysing the root-mean-square error it is seen that, despite the average differences, all correlations studied fit fairly well the observed data with values of $RMSE$ ranging from 3.2 to 3.9. Also, the obtained percentage errors remain below the 20% limit assigned during the experimental design.

### B. Continuous chemical flame height as a function of the measured flame temperature

It is worth remembering that the increase of flame temperature is the result of poor or complete combustion. Likewise, to reiterate that complete combustion has been achieved, it is assumed that all reactions end with the generation of $H_2O$ and $CO_2$. This implies that there is a temperature associated with the final combustion products, which are no longer considered to be part of the flame height since their density is different. Yagi [61] established a maximum flame temperature of 1473 K for hydrocarbons where the $CO_2$ fraction is at its maximum value. This characteristic establishes a length located between the bright and flickering flame zone. This temperature value was taken as a reference to locate a flame height for this arrangement of burner nozzles.

To determine the temperature, 32 measurements with 6 s exposure using a pyrometer with a Type C thermocouple probe [62] were performed for each fuel flow. Given the invasive nature of the probe as well as its sensitivity, no stratified measuring was conducted. Thus, measurements were taken only at half the average height along the centreline of the observable luminous flame body by employing a single-column mechanical positioner. The results are shown in Fig. 11.

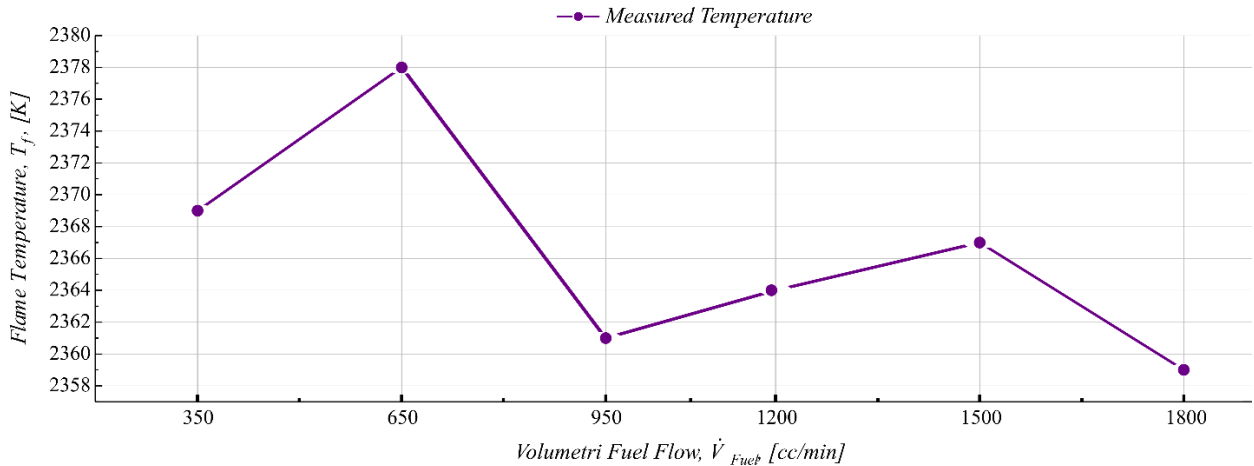

Fig. 11: Average measurements of maximum flame temperature, $T_{Flame}$, as a function of the fuel flow.

For this particular instance, the resulting theoretical magnitude equals to ~2398 K which agrees with the values obtained by Silverman [47]. From Fig. 11 it follows that the measured experimental temperature of the flame presents a relatively constant behaviour, despite the changes on the fuel flow, with a coefficient of variation of 0.29% and an overall average of 2366 K. Furthermore, the average discrepancy between theoretical and experimental magnitudes lies within 20 K, which is not statistically significant given that $P_{(F)} = 0.2739 > \alpha = 0.05$.

A complementary evaluation is also presented in this study, which consisted on performing temperature-based measurements of the flame. This test was devised as a mean to measure the flame length based not on the observable luminous reaction, but instead on the temperature distribution of chemical reactions during combustion. This will allow us to quantify and compare the length difference associated to both approaches. These measurements are summarized in Table IX. By means of thermal imaging, 32 images of the flame body were captured per fuel flow. A sample of the resulting images is displayed in Fig. 12.

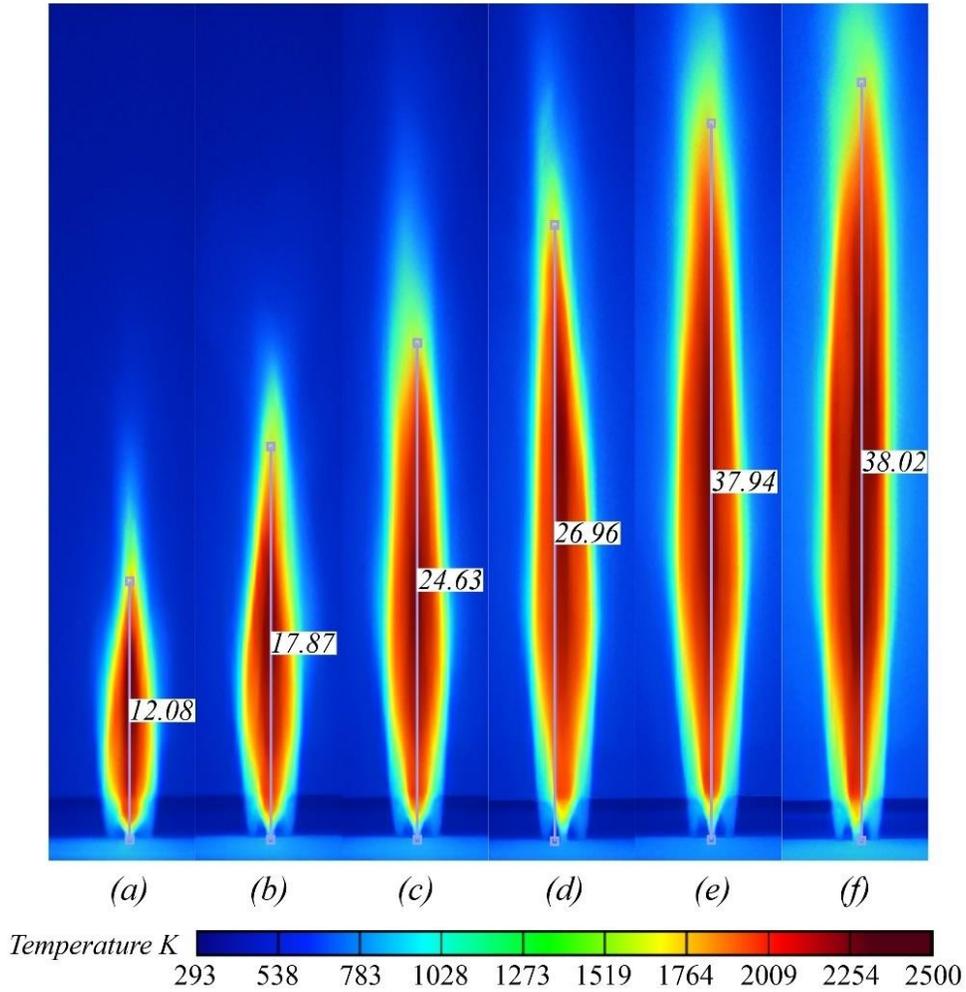

Fig. 12: Sample of the output measurements of the thermal-imaging camera.
(a) 350 cc/min, (b) 650 cc/min, (c) 950 cc/min, (d) 1200 cc/min, (e) 1500 cc/min and (f) 1800 cc/min.

Table IX: Main results/statistics of the chemical flame height measurements.

| Image source | Volumetric fuel flow | Replications | Mean flame height | Standard deviation | Variance | Range | Coeff. Of variation |
|---|---|---|---|---|---|---|---|
| Thermal Imaging | 350 | 32 | 11.571 | 0.893 | 0.799 | 3.982 | 7.73 |
| | 650 | 32 | 19.328 | 1.409 | 1.987 | 5.681 | 7.29 |
| | 950 | 32 | 23.909 | 1.095 | 1.199 | 4.855 | 4.58 |
| | 1200 | 32 | 29.571 | 2.832 | 8.020 | 10.392 | 9.58 |
| | 1500 | 32 | 34.993 | 1.856 | 3.445 | 7.081 | 5.30 |
| | 1800 | 32 | 37.139 | 2.093 | 4.380 | 8.574 | 5.64 |

In order to assess the effect of the thermal imaging employed on the flame height measurement, a similar analysis to the one in the previous section was performed. The results of this analysis are condensed in Tables X-XIV.

Table X: Two-way ANOVA for the setup comparison.

| Source | Sum of squares | Degrees of freedom | Mean square | F-Ratio | P-Value |
|---|---|---|---|---|---|
| *Main effects* | | | | | |
| R:Camera | 441.035 | 1 | 441.0355 | 145.42 | ~0.000 |
| C:Volumetric Flow | 29932.272 | 5 | 5986.454 | 1973.83 | ~0.000 |
| *Interactions* | | | | | |
| I:RC | 174.0578 | 5 | 34.811 | 11.48 | ~0.000 |
| *Residual* | | | | | |
| Error | 1128.244 | 372 | 3.0329 | | |
| Total | 31675.610 | 383 | | | |

Based on the *P*-Values, $P_{(F)}$, all the sources exert an effect on the output, with the fuel flow as the highest contributor to the variation of the measurements (see Table X). However, for this particular evaluation, within a 95% CI the use of a thermal-imaging camera has a statistically significant effect on the output.

As seen from the *F* statistic, the values for the camera and interaction sources are higher than the ones obtained in the previous analysis, implying that greater differences arise between the use of a high-definition and thermal-imaging camera, as well as their interaction with the changes on the fuel flow. These are attributed to the fact that the thermal imaging captures the heat release by the chemical reactions. Even though the observable luminous flame may undergo disturbances or instabilities, the length of the chemical reaction zone remains relatively more stable.

Once it is proved that changing the camera affects the measurement, it will be necessary to determine whether the output data comes from a normal distribution in order to conduct a proper error-uncertainty analysis. The results of the distribution fitting are shown in Table XI. Based on the values obtained, the possibility that the output data comes from a normal distribution cannot be rejected within a 95% CI, with all the calculated *P*-values, $P_{(SW)}$ and $P_{(KS)}$, being higher than the significance level previously assigned.

Table XI: Test for normality of the output values.

| Image source | Volumetric fuel flow | Shapiro-Wilk P-Value | Kolmogorov-Smirnov P-Value |
|---|---|---|---|
| Thermal Imaging | 350 | 0.759 | 0.957 |
| | 650 | 0.558 | 0.834 |
| | 950 | 0.877 | 0.993 |
| | 1200 | 0.657 | 0.968 |
| | 1500 | 0.124 | 0.659 |
| | 1800 | 0.658 | 0.795 |

The magnitude of the variation between the two approaches can be estimated by calculating the statistical tolerance from the observed measurements, which resulted to be $RSS_{SE} = 0.787$ cm, which is 38% lower than the 1.27 cm value obtained with the HD camera. This value indicates that the approach yields even better results with an overall error (that can be present during the treatments) ranging from 9.89% to 3.28% of the measurement, depending on the fuel flow studied (see Table XII).

Table XII: Standard error of the measurement and statistical tolerance.

| Image source | Volumetric fuel flow | Standard error | Root sum of squares |
|---|---|---|---|
| Thermal Imaging | 350 | 0.158 | 0.787 |
| | 650 | 0.249 | |
| | 950 | 0.193 | |
| | 1200 | 0.5006 | |
| | 1500 | 0.328 | |
| | 1800 | 0.37 | |

As shown in Table XIII an expanded uncertainty of $U = \pm 0.642$ cm is achieved, which is slightly higher than the one obtained for the HD-camera approach. This is attributed to the intrinsic variations of the algorithm developed for measurement, given that it was not originally tailored to properly identify and crop the colour map associated to thermal imaging. Similarly, given that, $U \approx RSS_{SE}$, it is concluded that most of the measurement error is attributed to the uncertainty of the approach. Nonetheless, based on the overall results and low magnitudes, it is concluded that the output of the algorithm is valid and acceptable.

Table XIII: Type A uncertainty measurement results.

| Pooled variance | Pooled standard deviation | Standard uncertainty | Coverage factor | Expanded uncertainty | Relative uncertainty |
|---|---|---|---|---|---|
| 2.875 | 1.818 | 0.321 | 2 | 0.642 | 7.76% |

This conclusion is further confirmed by the error values reported in Table XIV. Given the validity of the previously obtained luminous flame heights, the discrepancy between the approaches could be considered to be an error, thereby corroborating the previously assumed approximate lengthwise difference of 2 cm between the luminous and the heat release zone due to the chemical reactions. The average difference between measurements amounts to 2.331 cm, while the root-mean-square error yields 2.539 cm. Since these two values are rather equal, it is concluded that the variation in the differences between the approaches is practically null and that the magnitude of such differences is not particularly large. In addition, it is confirmed that most of the intrinsic bias related to the experiment conducted in the previous section is indeed product of the randomness of the combustion process and the particularities of the resulting observable flame structure.

Table XIV: Assessment of the differences due to the use of thermal imagery.

| Image source | Mean absolute error | Root mean square error | Mean absolute percentage error |
|---|---|---|---|
| Thermal Imaging | 2.331 | 2.539 | 9.983 |

It is worth mentioning that the temperature-based assessment, presented in this section, requires a lengthier experimental design and procedure as well as further statistical analysis. Nonetheless, based on the preliminary results and the information obtained from them, it is considered that the difference between the visualization devices is not particularly large. In fact, it lies within the expected limits of the proposed design and is attributable to known phenomena within the combustion process. Therefore, the methodology presented in this paper is suitable for flame length measurements and is proposed as an alternative methodology to the existing ones.

## C.     Empirical correlation for laminar-transition diffusion flame height measurement

From the height and temperature magnitudes obtained through the experimental procedure described above, it is inferred that a correlation exists between the studied factors. For the measured flame heights, it is found that the fuel flow, $\dot{V}_{Fuel}$, is the measured variable with higher significance, while the effect of the flame temperature is practically negligible, having *P*-values of ~0.00005 and 6.41, respectively. This output confirms the qualitative estimation given in Section II, leading to the initial assumption that the height of the flame body could be determined solely through the magnitude of the volumetric fuel flow and the associated flow regime.

Based on the analysis of the output datasets, statistical modelling is performed and correlations for the laminar-transition diffusion flame height measurement are derived as

$$h_{cf}\big|_x = (0.0125 + 26.10/\dot{V}_{Fuel})^{-1}, \tag{23}$$

$$h_{if}|_x = (0.1994 + 203.4274/\dot{V}_{Fuel})^{-1}. \tag{24}$$

Equations (23) and (24) are obtained from a double reciprocal regression, i.e., $y = (a + b/x)^{-1}$, and their combined behaviour is displayed in Fig. 14 as compared to the employed benchmarks.

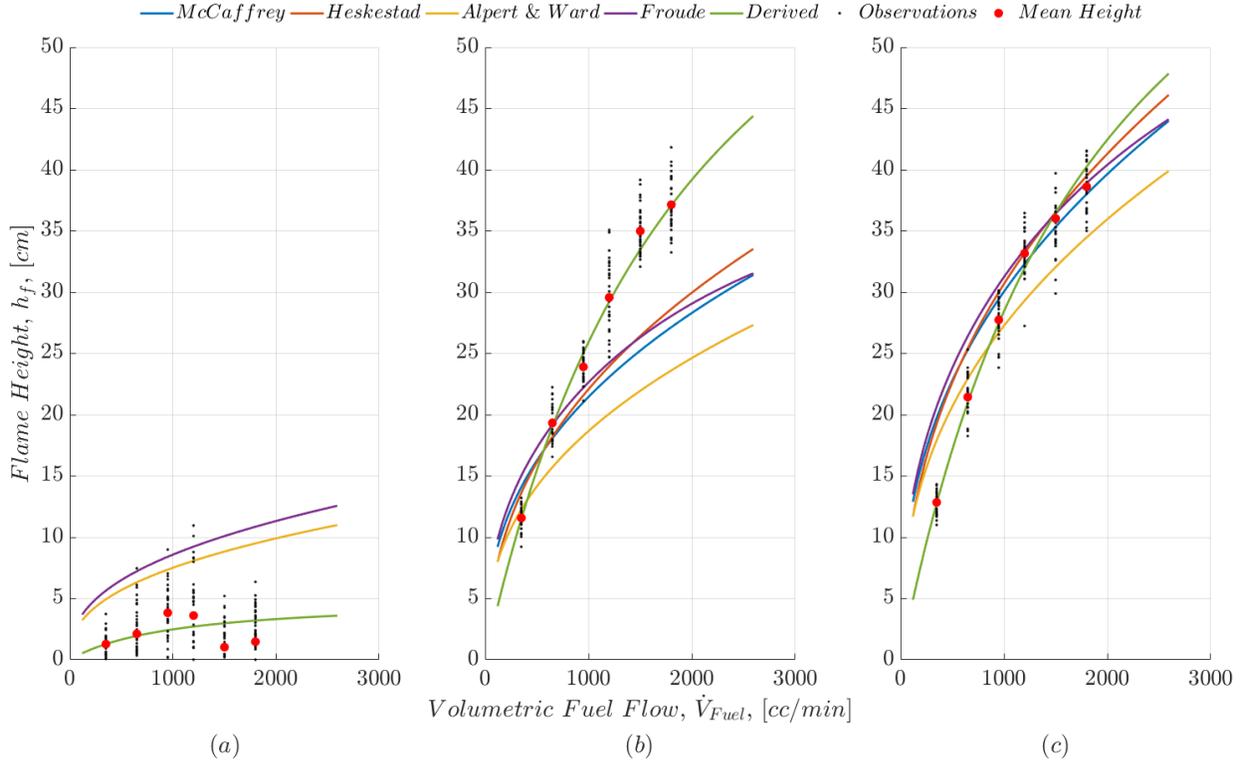

(a)  (b)  (c)

Fig. 13: Behaviour of the derived correlation for laminar-transition flame height.
(a) Intermittent, (b) continuous and (c) luminous

It is seen from Fig. 13 that the derived correlation exhibits a rather steady behaviour up to a certain point, with the main difference that the intermittent-flame height being persistently lower and the continuous-flame height being higher, as shown in Figs. 13a and 13b, respectively. This discrepancy is attributed to the nature of the analysed flame, specifically, a jet fire. Compared to a pool fire, it presents a higher flame height based on the fact that the fuel is being supplied continuously in gaseous form and with a significantly higher momentum in upward direction, leading to a much higher heat flux and therefore to a larger flame. This fuel delivery mechanism also explains the lower intermittent height, since both are correlated. A larger continuous flame reduces the flame intermittency, confining it mainly to the tip of the flame particularly at high Reynolds numbers. This correlation and magnitude could be refined further by employing a high-speed camera. This will account for the frequency of these phenomena and thoroughly define a more accurate length where flame flickering/puffing is expected to occur.

Finally in Fig. 13c the luminous flame height correlation is depicted by just adding the magnitudes depicted in Figs. 13a and 13b. This correlation has a coefficient of determination, $R^2$, of 97.23%, which explains the percentage variability of the flame height based on the fitted model; a mean absolute error of 0.0022cm as the average value of the residuals and a standard error of estimates of 0.0038cm, describing the negligible deviation of such residuals. These values are considered more than acceptable, in terms of model adjustment. Thus, the proposed correlation could be used to construct prediction limits for new observations.

# CONCLUSIONS

This paper presents a thorough experimental methodology for luminous flame height measurement of a 4-nozzle port-array burner, which is based on a comprehensive experimental design and statistical analysis as a means for indirect measurement, mainly based on convolution operations for edge detection and digital mapping for pixel quantification. Furthermore, the experimental observations were compared against theoretical forecasts to determine how well they fit the behaviour described by empirical correlations in the literature. As a complement to the present study, an evaluation was also made that consisted of carrying out flame measurements based on the temperature distribution of chemical reactions during the combustion process using thermal images. This has made it possible to quantify and compare the flame length difference associated with observations using high definition cameras and a thermographic camera.

It was observed that the lateral camera presented an average $CC_V$ of 7.78%, the front camera of 5.93% and the angled camera of 5.67%. This was because the angled camera was able to capture characteristics of the flame structure, such as flame tilt, tip flicker and/or flame kernels still attached to the front of the flame in comparison to the remaining planes.

A relatively low $F$ statistic of the camera direction implies that, although the camera affects the measurements, flame height variations between the planes are minimal compared to the fuel flow. When analysing the $F$ statistic of the volumetric fuel flow it was revealed that this is the factor that contributes the most to the variability of the flame structure since the higher the fuel load, the more intense the flame becomes. These findings reveal that the intrinsic error associated with each camera is relatively low, suggesting that most of the variability should be attributed solely to flame distortion, flame flickering and the occasional capture of still-attached kernels to the flame front. Based on the error-uncertainty relation of $2U \approx RSS_{SE}$, both the experimental design and the procedure presented in this document can be considered to be acceptable, implying that the observations tend to remain relatively centred and that the uncertainty measure obtained does not favour any HD camera direction in particular. This in turn suggests that the systematic error did indeed become a random error, thus confirming why the persistent bias is related to the random behaviour of the combustion process.

It was also found that the observed experimental data fit the theoretical models for the measurement of the flame height. In particular, this is so for the larger fuel flow rates 950, 1200, 1500 and 1800 cc/min, where the maximum relative error amounts to 10% for the 950 cc/min fuel flow. Alternatively, for the remaining fuel flows, the maximum relative error increases to 23% and 60%, for the 650 cc/min and 350 cc/min cases, respectively. This is mainly because diffusion flames at low Reynolds numbers tend to become unstable under the effect of buoyancy-induced vortices, causing massive and difficult to measure flickering.

When using the thermographic camera, it was observed that it has a statistically significant effect. This is attributed to the fact that thermal image captures the heat released by chemical reactions. Although the observable luminous flame may exhibit disturbances or instabilities, the length of the chemical reaction zone remains relatively more stable. Since $RSS_{SE} = 0.787$ cm is 38% smaller than the 1.27 cm value obtained with the HD camera, the thermographic approach gives even better results, leading to a general error ranging from 9.89% to 3.28% of the measurement, depending on the fuel flow. The average difference between the measurements amounts to 2,331 cm, while the mean square root error is 2,539 cm. Given that these two values are similar, the variation in the differences between the approaches is practically null and so the magnitude of these differences is not particularly large. Furthermore, it is assessed that most of the intrinsic bias related to the experiment is in fact a product of the randomness of the combustion process. Particularly at high Reynolds numbers a larger continuous flame develops reducing the intermittency and confining it mainly to the tip of the flame.

The resulting measures fit quite well all the proposed empirical correlations, with a root mean square error that varies from 3.247 to 3.968. In contrast, based on the lower absolute mean errors, a better fit is achieved when the McCaffrey and Alpert & Ward correlations are used because they handle flame flicker much better. However, given that all the correlations have an average absolute percentage error of less than 20%, the resulting measures represent an acceptable fit compared to the available benchmarks. Thus, the proposed correlations explain, on average, 97.23% of the variability of the height of the measured luminous flame.

Therefore, the methodology based on image convolution and statistical validation for diffusion flame heights estimation in laminar to transition-to-turbulent regimes presented in this work can be considered to be appropriate and can be proposed as an alternative methodology to the existing ones. Finally, this information can be used to optimize and improve the efficiency of diffusion flames in the combustion processes of current multi-port burner technologies.

## ACKNOWLEDGEMENTS

The authors acknowledge the support provided by the Institute of Engineering of the National Autonomous University of Mexico and grants provided by the National Council of Science and Technology of Mexico (CONACyT) and the Mexican Ministry of Energy (SENER). The support provided by the Laboratory of Applied Thermal and Hydraulic Engineering, Superior School of Mechanical and Electrical Engineering of the National Polytechnic Institute is also acknowledged.

## AUTHOR CONTRIBUITIONS

**M. De La Cruz-Ávila**: Conceptualization, Investigation, Data Curation, Resources, Visualization, Writing – Original Draft, Review & Editing, Project Administration, Funding Acquisition;
**J.E. De León-Ruiz**: Conceptualization, Methodology, Software, Validation, Formal Analysis, Investigation, Data Curation, Visualization, Writing – Original Draft, Review & Editing;
**I. Carvajal-Mariscal**: Methodology, Software, Validation, Formal Analysis, Investigation, Resources, Writing – Review & Editing, Supervision, Project Administration;
**F. Peña-Polo**: Methodology, Validation, Visualization, Writing – Review & Editing.
**L. Di G. Sigalotti**: Validation, Writing – Review & Editing.